\documentclass[pmlr,twocolumn,10pt]{jmlr} 

\mlhtrack{proceedings}

\newif\iffinal
\finaltrue  

\iffinal
    \ifmlhneedspmlr
      \jmlrvolume{346}
      \jmlryear{2026}
    \fi
    \ifmlhfindings \jmlrproceedings{}{ML4H 2026 - Findings Track}\fi
    \ifmlhdemo     \jmlrproceedings{}{ML4H 2026 - Demo Track}\fi
    \jmlrworkshop{Machine Learning for Health (ML4H) 2026}
\else
    \jmlrproceedings{}{Submitted to ML4H 2026: \mlhtrackname}
    \jmlrworkshop{Machine Learning for Health (ML4H) 2026}
\fi

\usepackage{booktabs}
\usepackage{siunitx}

\usepackage[switch]{lineno}

\theorembodyfont{\upshape}
\theoremheaderfont{\scshape}
\theorempostheader{:}
\theoremsep{\newline}

\title[Short Title]{Mammography Foundation Models for Opportunistic MACE Prediction}

\title{Mammography Foundation Models for Opportunistic Prediction of Major Adverse Cardiovascular Events}

\author[Feldman et al.]{
\Name{Paula Feldman\textsuperscript{1,3*}} \Email{paf4004@med.cornell.edu}\\
\Name{Nusrat Binta Nizam\textsuperscript{2, 3}}\Email{nn284@cornell.edu}\\
\Name{Sunwoo Kwak\textsuperscript{2, 3}}\Email{sk3355@cornell.edu}\\
\Name{Batuhan Karaman\textsuperscript{2, 3}}\Email{kbk46@cornell.edu}\\
\Name{Katerina Dodelzon\textsuperscript{2}}\Email{kad9090@med.cornell.edu}\\
\Name{Mert R. Sabuncu\textsuperscript{1,2,3}}\Email{msabuncu@cornell.edu}\\
\addr
\textsuperscript{1} Department of Radiology, Weill Cornell Medicine, New York, NY, USA\\
\textsuperscript{2} Cornell University, New York, NY, USA\\
\textsuperscript{3} Cornell Tech, New York, NY, USA
\medskip
\noindent
\\\textsuperscript{*}Corresponding author.}

\begin{document}

\maketitle

\ifmlhdemo\else
\begin{abstract}
Cardiovascular disease (CVD) remains the leading cause of death among women, yet cardiovascular risk assessment often relies on clinical variables that may be missing, outdated, or unavailable in routine care. Screening mammography offers an opportunity for opportunistic cardiovascular risk stratification because it is routinely acquired and contains vascular features, including breast arterial calcifications (BAC), that are associated with cardiovascular risk and events. We evaluate whether mammography-specific foundation models, originally pretrained for breast cancer-related tasks, can transfer to cardiovascular risk prediction without cardiovascular-specific supervision or explicit BAC annotation. We constructed a 5-year major adverse cardiovascular event (MACE) cohort of 22,497 women linked to electronic health record outcomes, including 500 events (2.22\% prevalence).Using only the screening mammogram, the foundation models achieved AUROCs of 0.823 and 0.822, recovering most of the gain of a clinical risk-factor model (AUROC 0.859) over an age-only model (0.765), without requiring any clinical variables. Both foundation models evaluated assigned substantially higher predicted risk to patients with radiologist-documented BAC, despite BAC never being used as a training label, and showed activation patterns consistent with vascular findings. Together, these findings suggest that mammography foundation models can recover clinically relevant cardiovascular risk information directly from mammographic pixels and suggest that screening mammography may provide an opportunistic source of cardiovascular risk information to complement conventional clinical assessment without additional imaging. 

\end{abstract}
\fi

\ifmlhneedsstatements
\paragraph*{Data and Code Availability}
The clinical data used in this study are not publicly available due to institutional and patient privacy restrictions.  Code is available in https://github.com/PauFeld/MammoCVD

\paragraph*{Institutional Review Board (IRB)}
This retrospective study was approved by the Institutional Review Board (IRB) of Weill Cornell Medicine (Protocol No. 22-08025152). All data were de-identified prior to analysis in accordance with institutional policies and applicable privacy regulations. No directly identifiable patient information was accessed or used in this study.
\fi

\section{Introduction}
\label{sec:intro}

Cardiovascular disease (CVD) remains the leading cause of death among women worldwide, yet primary prevention relies heavily on structured clinical variables \citep{martin20252025, arnett2019primary, mikhail2005coronary} that may be missing, outdated, or unavailable at the time of routine clinical encounters, limiting opportunities for timely identification of women who may benefit from preventive intervention. Furthermore, established risk prediction models used in clinical practice, such as the Predicting Risk of Cardiovascular Disease Events (PREVENT) calculator, may underestimate risk in some women, particularly those with atypical presentations~\citep{zhou2025evaluation, khan2023novel, baart2019cardiovascular}.

Screening mammography offers a potential opportunity to address this gap. Mammography is routinely performed in a large proportion of women, including many who have not undergone dedicated cardiovascular evaluation. Importantly, breast arterial calcifications (BAC) are readily visible on mammography and have been associated with cardiovascular disease and subsequent cardiovascular events~\citep{margolies2026breast}. BAC, once considered an incidental mammographic finding, is now recognized as a potential marker of cardiovascular risk, with FDA-cleared tools such as cmAngio (CureMetrix, Inc., San Diego, CA) now available for automated detection. Early cohort studies established associations between BAC and cardiovascular mortality and incident coronary disease \citep{kemmeren1998arterial, iribarren2004breast}, while subsequent studies extended these findings to stroke, heart failure, and broader outcomes \citep{loberant2013prevalence, iribarren2022breast}. More recent work has moved beyond visual assessment toward automated quantification of BAC, demonstrating its potential as a quantitative cardiovascular risk biomarker \citep{dapamede2026bac}.

However, most existing approaches rely on the explicit detection or quantification of BAC or other predefined imaging phenotypes~\citep{nerlekar2026novel, watanabe2025artificial, guo2021scu}. While these studies establish the feasibility of extracting cardiovascular risk information from mammographic images, they leave open whether such signal can be recovered without cardiovascular-specific supervision or explicit identification of a predefined biomarker. Moreover, requiring BAC assessment adds time and complexity to the radiologist's workflow. We hypothesize that cardiovascular risk-related signal can instead be learned directly from mammographic pixels. This is particularly relevant in the era of medical imaging foundation models, which can learn rich representations from large, domain-specific datasets and subsequently be adapted to downstream clinical tasks\citep{van2025foundation, yang2025chest, chen2019med3d}. In mammography, recent foundation models such as Mammo-CLIP and Mammo-FM have been pretrained on large collections of mammograms and associated clinical or textual information for breast cancer-related tasks \citep{ghosh2024mammoclip, ghosh2025mammofm}. Their learned representations may therefore capture information extending beyond the specific tasks used during pretraining. Whether such representations can transfer to clinically distinct domains, however, remains largely unexplored.

In this work, we evaluate whether mammography-specific foundation models pretrained for breast cancer-related tasks can be transferred to cardiovascular risk prediction. Specifically, we evaluate 5-year major adverse cardiovascular events (MACE) prediction from screening mammograms linked to electronic health record outcomes. Importantly, these foundation models were not pretrained using cardiovascular outcomes and were not explicitly trained to identify BAC or any other cardiovascular imaging biomarker. We therefore ask whether cardiovascular risk information can be recovered directly from representations learned for a different clinical task, without cardiovascular-specific pretraining or expert annotations.

\textbf{Contributions.} (1) We construct a cohort of women undergoing screening mammography with linked electronic health record outcomes. (2) We assess whether mammography foundation models pretrained on breast cancer tasks transfer to cardiovascular risk prediction without cardiovascular-specific pretraining, BAC annotation, or handcrafted imaging features. (3) We benchmark image-based predictions against clinical risk-factor and age-only baselines at clinically motivated operating points. (4) We use interpretability analyses to examine the imaging signal underlying predictions and its association with recognizable vascular features. Together, these findings provide evidence that mammography foundation models can capture information relevant to cardiovascular risk and suggest the potential for opportunistic screening from routinely acquired mammograms.

\begin{figure*}[t]
  \centering
  \includegraphics[width=\textwidth]{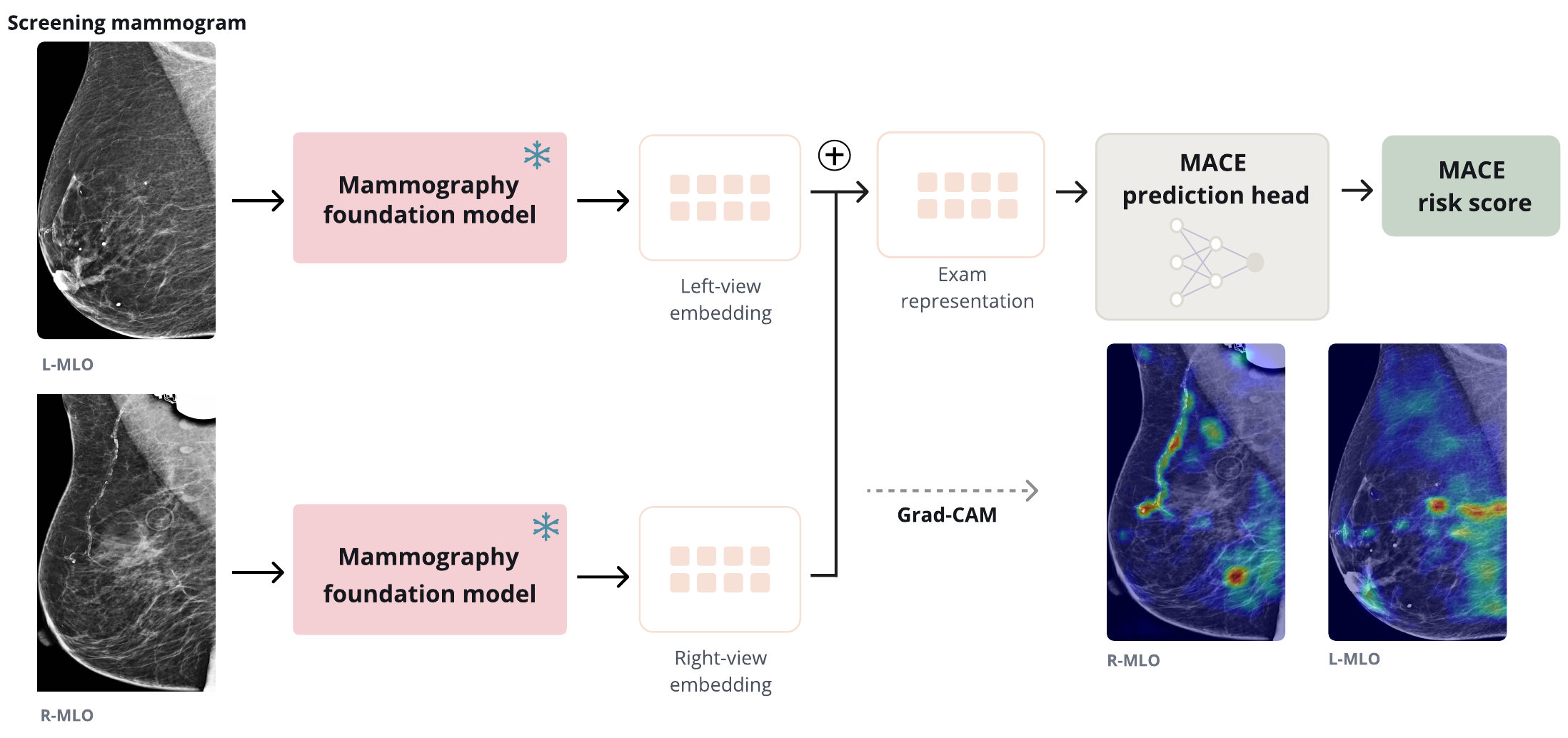}
  \caption{Overview of the proposed transfer-learning framework. A
  mammography foundation model pretrained for a breast-imaging task is adapted
  to predict 5-year cardiovascular risk from screening mammograms. }
  \label{fig:architecture}
\end{figure*}

\section{Methods}
\label{sec:methods}

\textbf{\subsection{Cohort construction}}
\label{sec:cohort}

Our study included 28,746 women who underwent screening mammography at Weill Cornell Medicine between 2014 and 2019. Screening mammograms were linked to structured electronic health record (EHR) diagnosis and laboratory data through institutional OMOP-formatted tables.
The prediction target was a fixed 5-year major adverse cardiovascular event (MACE) composite comprising acute myocardial infarction, ischemic stroke, or acute heart failure, identified using a predefined set of ICD-9/10 codes. Patients with a qualifying event before the index mammogram were excluded, as were those with an event within 180 days of the index mammogram to reduce potential label leakage. Patients were labeled positive if a qualifying event occurred within 5 years after the index mammogram and negative if they remained event-free for at least 5 years. Patients censored before completing 5 years without an event were excluded.

After exclusions due to insuficient follow up, the final cohort comprised 22,497 patients with at least one screening mammogram and a determinable 5-year MACE outcome, including 500 patients with a qualifying event (2.22\%). MACE components were not mutually exclusive; 18 patients had multiple components recorded on the qualifying event date. Following prior approaches, only mediolateral oblique (MLO) views were used because they provide greater visualization of the relevant vascular regions.

Patients were split at the patient level into a fixed held-out test set (5,287 patients, 23.5\%; 118 events) and a remaining development cohort (17,210 patients; 382 events) for 5-fold cross-validation. Within each fold, training and validation sets were stratified by MACE outcome, comprising approximately 61.2\% and 15.3\% of the full cohort, respectively. The held-out test and validation sets used a single scan per patient (the first qualifying screening mammogram). The training set was expanded through multi-instance augmentation: each available scan from a training patient was independently labeled as MACE positive or negative relative to its acquisition date. A scan was retained as a positive instance if a qualifying MACE occurred within the subsequent 5 years, as a negative instance if the patient had at least 5 years of subsequent event-free follow-up, and discarded otherwise. Table~\ref{tab:clinical-vars} summarizes the availability and distribution of clinical variables available. Continuous variables are reported as mean $\pm$ SD and categorical variables as count (\%) among patients with available data.

\begin{table}[t]
\centering
\caption{Characteristics of the study cohort.}
\label{tab:dataset}
\scriptsize
\begin{tabular}{lr}
\toprule
Characteristic & Overall \\
\midrule
Patients & 22,497 \\
MACE events & 500 (2.22\%) \\
\midrule
Acute myocardial infarction & 113 \\
Ischemic stroke & 253 \\
Acute heart failure & 191 \\
\midrule
Age, mean (SD) & 55.7 (11.9) \\
Dense breast (ACR C/D) & 64.0\% \\
\bottomrule
\end{tabular}
\end{table}

\subsection{Downstream prediction task}
\label{sec:task}

The primary task is binary prediction of 5-year MACE from a single 
screening mammogram. The model receives both MLO mammographic views from
the index examination and predicts whether the patient will experience a
qualifying MACE within the subsequent 5 years. The prediction horizon is
fixed across all patients, and no information occurring after the index
examination is provided to the image-based prediction models.

\textbf{Outcome definition.}
The MACE composite comprised non-fatal acute myocardial infarction, non-fatal ischemic stroke, or acute heart failure, identified using a predefined set of ICD-9/10 diagnosis codes. Acute myocardial infarction was defined by ICD-10 I21 (all subcodes) or ICD-9 410 (all subcodes). Ischemic stroke was defined by ICD-10 I63 (all subcodes) or ICD-9 infarction-specific subcodes (433.01, 433.11, 433.21, 433.31, 433.81, 433.91, 434.01, 434.11, 434.91); hemorrhagic and ill-defined stroke codes (I60, I61) were excluded. Acute heart failure was defined by ICD-10 codes for acute or acute-on-chronic decompensation (I50.813, I50.43, I50.33, I50.23, I50.21) and corresponding ICD-9 codes (428.43, 428.33, 428.23, 428.41). MACE components were not mutually exclusive; 18 of 500 positive patients had multiple qualifying components recorded.

\begin{table}[t]
\centering
\caption{Clinical variables in the cohort based on the first available mammogram for each patient. Continuous variables are reported as mean $\pm$ SD and categorical variables as count (\%) of the full cohort.}
\label{tab:clinical-vars}
\scriptsize
\begin{tabular}{lcc}
\toprule
Variable & Available & Summary \\
\midrule
\multicolumn{3}{l}{\textit{Continuous}} \\
Age (years) & 100.0\% & 55.7 $\pm$ 11.9 \\
Total cholesterol (mg/dL) & 65.7\% & 196.5 $\pm$ 34.1 \\
LDL (mg/dL) & 61.9\% & 112.5 $\pm$ 28.6 \\
HDL (mg/dL) & 62.5\% & 64.5 $\pm$ 17.2 \\
Triglycerides (mg/dL) & 62.1\% & 97.2 $\pm$ 56.8 \\
Creatinine (mg/dL) & 75.8\% & 0.8 $\pm$ 0.4 \\
HbA1c (\%) & 35.5\% & 5.8 $\pm$ 0.8 \\
\midrule
\multicolumn{3}{l}{\textit{Categorical}} \\
Diabetes & 100.0\% & 2,645 (11.8\%) \\
Hypertension & 100.0\% & 5,088 (22.6\%) \\
Current/former smoker & 85.0\% & 1,583 (7.0\%) \\
Alcohol use & 76.1\% & 10,712 (47.6\%) \\
Family history of CVD$^{a}$ & 58.9\% & 13,254 (58.9\%) \\
\bottomrule
\end{tabular}
\vspace{2pt}
\begin{minipage}{\linewidth}
\footnotesize
\raggedright
$^{a}$ Family history of CVD is a presence-only EHR flag; the count therefore represents patients with a positive flag.
\end{minipage}
\end{table}

\subsection{Mammography foundation-model transfer}
\label{sec:transfer}

We evaluated two publicly released mammography foundation models, Mammo-CLIP \citep{ghosh2024mammoclip} and Mammo-FM \citep{ghosh2025mammofm}. Both models use EfficientNet-B5 image encoders~\cite{pmlr-v97-tan19a} to map individual mammographic views to 2,048-dimensional embeddings. Mammo-CLIP uses a CLIP-style image--report contrastive framework, while Mammo-FM extends this approach with image--report, image--image, and text--text contrastive objectives. We used the pretrained, frozen image encoders to extract embeddings from each MLO view (L-MLO and R-MLO). Because neither model includes explicit multi-view fusion, we mean-pooled the two view embeddings to obtain a single exam-level representation. A two-layer multilayer perceptron (MLP) was then trained on this representation to predict 5-year MACE risk, with gradients propagated only through the MLP risk head. 

Mammo-CLIP was pretrained on 13,829 patient--report pairs from the University of Pittsburgh Medical Center (UPMC), corresponding to 25,355 screening mammograms restricted to BI-RADS 0--2 examinations \citep{ghosh2024mammoclip}. Mammo-FM was pretrained on a substantially larger, multi-institutional corpus of approximately 140,000 patients and 800,000 mammograms from UPMC, EMBED, Boston Medical Center, and Mayo Clinic \citep{ghosh2025mammofm}. Thus, the models differ in pretraining scale, institutional diversity, and contrastive objectives. Neither model was pretrained on cardiovascular outcomes or BAC-specific labels.

\subsection{Statistical analysis.} For each of the five cross-validation folds, a model was trained on the corresponding training set, with early stopping and model selection performed exclusively on the validation set. This procedure was repeated using three random seeds, yielding 15 models in total. All models were evaluated on the same held-out test set. Predictions were first averaged across seeds within each fold and then across folds, producing a single ensemble risk score for each test patient. All reported test-set performance metrics are based on these ensemble predictions.
Separately, the out-of-fold validation predictions, comprising one prediction per patient across the development cohort, were pooled and used exclusively to fit calibration models.

Uncertainty was estimated using nonparametric bootstrap resampling of held-out test patients (2,000 resamples). We report 95\% percentile confidence intervals. Model comparisons were paired: for each bootstrap resample, all models were evaluated on the same patients, and differences in performance metrics were calculated. We report the observed difference, its 95\% percentile confidence interval, and a two-sided bootstrap $p$-value.

\subsection{Implementation details}
\label{sec:training}

For probing, the pretrained image encoder is kept strictly frozen, and only the multilayer perceptron is trained on the downstream MACE prediction task.
The probe is optimized using binary cross-entropy with positive-class weighting to account for the low event prevalence. Optimization uses AdamW with a learning rate of $1\times10^{-4}$ and weight decay of $5\times10^{-4}$. Training uses a batch size of 64, gradient clipping with maximum norm 1.0, and a maximum of 50 epochs. Early stopping is based on validation AUROC with a patience of 8 epochs, and the checkpoint with the highest validation AUROC is selected for a single evaluation on the held-out test set.

\section{Evaluation and Results}
\label{sec:results}

\subsection{Discrimination}
\label{sec:discrimination}

We evaluated the ability of each model to discriminate patients who experienced a 5-year MACE event from those who did not. All models were evaluated on the same held-out test set of 5,287 patients, including 118 patients with a qualifying event. As baselines, we report a tabular model trained on clinical features, and the same model trained using age alone. The tabular model consists of a neural network using the variables listed in Tab.~\ref{tab:clinical-vars}, with values available on or before the mammogram date. Each variable is mapped to a learned 32-dimensional representation: continuous variables are linearly projected from their standardized values, while missing values are represented by a learned missing-value embedding. The variable representations are summed and passed to a two-layer MLP with a single output. 

Table~\ref{tab:main} summarizes test-set AUROC performance. The tabular clinical risk model achieved the highest discrimination (AUROC, 0.859), followed by Mammo-CLIP (0.823) and Mammo-FM (0.822), while the age-only baseline achieved an AUROC of 0.765. Both image-based models significantly outperformed the age-only baseline (paired bootstrap, $p=0.0016$ for Mammo-CLIP; $p=0.0040$ for Mammo-FM). Mammo-CLIP was not significantly different from the full clinical model ($p=0.057$), whereas Mammo-FM showed a small but statistically significant difference ($p=0.048$). Thus, Mammo-CLIP achieved discrimination comparable to the full clinical model using mammography alone, indicating that screening mammograms contain cardiovascular risk information beyond age.

\begin{figure}[t]
  \centering
  \includegraphics[width=\linewidth]{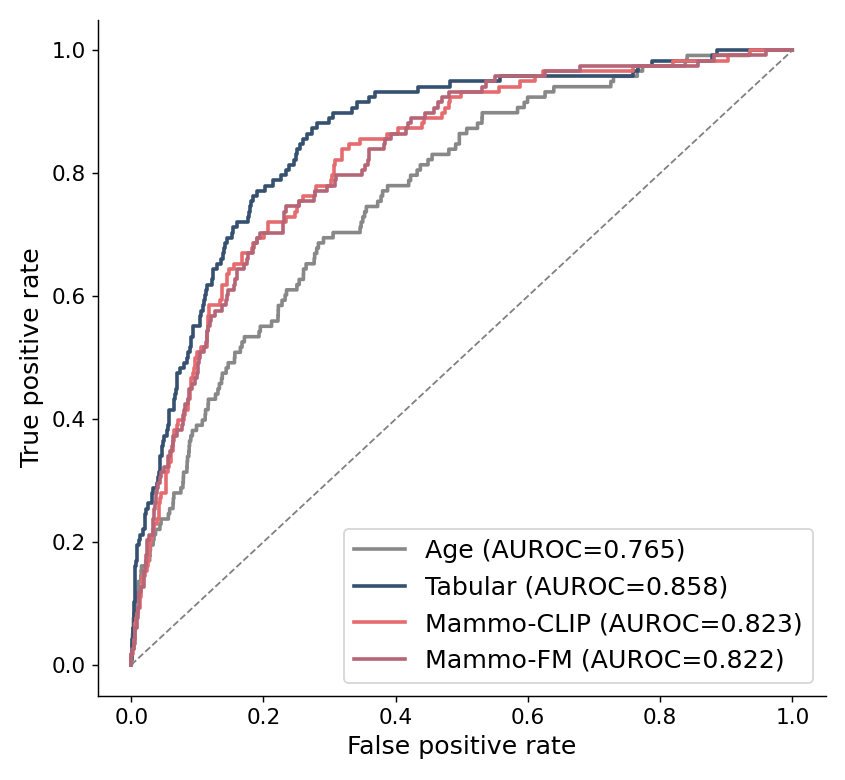}
  \caption{Receiver operating characteristic curves for prediction of
  5-year MACE on the held-out test set.}
  \label{fig:roc}
\end{figure}

\begin{table}[t]
  \centering
  \caption{Test-set discrimination for prediction of 5-year MACE.
  Confidence intervals are based on 2,000 bootstrap resamples.}
  \label{tab:main}
  \scriptsize
  \begin{tabular}{lcc}
    \toprule
    Model & AUROC & 95\% CI \\
    \midrule
    Age & 0.765 & [0.722, 0.805] \\
    Tabular & 0.858 & [0.825, 0.889] \\
    Mammo-CLIP & 0.823 & [0.785, 0.856] \\
    Mammo-FM & 0.822 & [0.784, 0.858] \\
    \bottomrule
  \end{tabular}
\end{table}

\paragraph{Effect of pretraining.} To assess the contribution of mammography-specific pretraining, we evaluated three additional image encoders using the same downstream pipeline as our primary models: frozen encoders, mean-pooled MLO views, and an MLP classification head. All experiments used the same cross-validation folds and random seeds. The encoders included (1) the same EfficientNet-B5 architecture initialized with ImageNet weights, (2) the image encoder from Mirai, a breast cancer risk prediction model \citep{yala2021toward}, and (3) a ResNet-18 pretrained using self-supervised DINO \citep{caron2021emerging} on screening mammograms from 28,665 patients at our institution, excluding test patients.

The ImageNet-pretrained EfficientNet-B5 achieved an AUROC of 0.815 (95\% CI, 0.781--0.847), comparable to Mammo-FM ($\Delta$AUROC = +0.007; 95\% CI, $-0.013$ to $+0.027$) and Mammo-CLIP ($\Delta$AUROC = +0.008; 95\% CI, $-0.013$ to $+0.030$). In contrast, Mirai and DINO achieved lower AUROCs of 0.777 (95\% CI, 0.736--0.817) and 0.756 (95\% CI, 0.717--0.794), respectively, significantly underperforming both mammography foundation models (all $p<0.001$).

\subsection{Risk enrichment among the highest-risk patients}
\label{sec:enrichment}

\begin{figure*}[t]
  \centering
  \includegraphics[width=\textwidth]{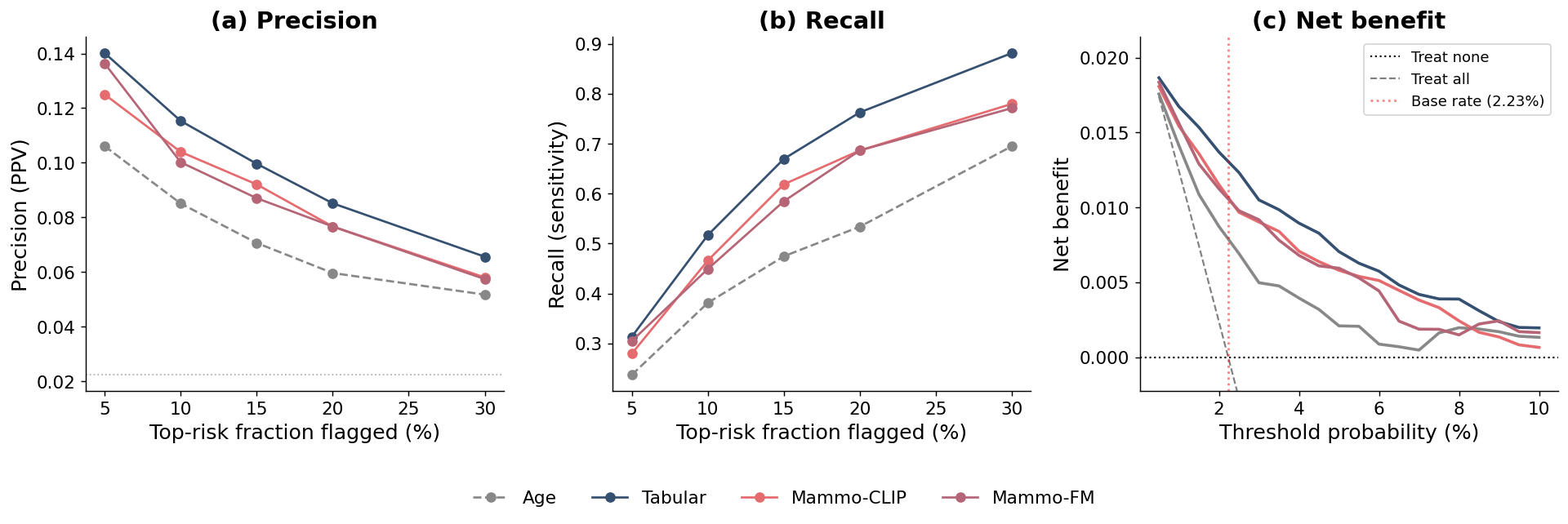}
  \caption{Test-set precision \textbf{(a)} and recall \textbf{(b)} among the top-risk-ranked $x\%$ of patients, and \textbf{(c)} net benefit for each model's predictions ($n=5{,}287$ test patients, including 118 with a 5-year MACE event). “Treat all” and “Treat none” reference strategies, along with the cohort event rate, are shown for comparison.
}
\label{fig:pr_netbenefit_combined}
  \label{fig:enrichment}
\end{figure*}

Because an opportunistic screening system would prioritize a manageable subset of patients for additional cardiovascular assessment, we evaluated how effectively each model concentrated future MACE events among patients at highest predicted risk. Figure~\ref{fig:pr_netbenefit_combined} shows precision (a) and recall (b) as progressively larger fractions of the test cohort were selected by predicted risk.

At the top 20\% of predicted risk Mammo-FM and Mammo-CLIP each captured 68.6\% of events with 7.7\% precision, corresponding to 3.4-fold enrichment. Across the evaluated risk fractions, both foundation models consistently outperformed the age-only baseline at the 5\%, 10\%, and 20\% thresholds.The foundation models' advantage over ImageNet appeared at high-risk operating points: when flagging the top 10\% of patients, Mammo-CLIP identified 46.6\% of events versus 38.1\% for ImageNet (+8.5 points; 95\% CI 0.7--17.0).

These results show that mammography-based foundation models can concentrate future cardiovascular events within a small, high-risk subset of the screening population using only the screening mammogram at inference, with performance approaching that of a clinical risk model based on established cardiovascular risk factors.

\subsection{Clinical utility and net benefit}
\label{sec:net_benefit}

To evaluate the potential clinical utility of opportunistic cardiovascular
risk prediction, we performed decision-curve analysis and estimated the
net benefit of each model across a range of risk thresholds. Net benefit was
calculated as
\[
\mathrm{NB}(p_t) =
\frac{TP}{N}
-
\frac{FP}{N}\frac{p_t}{1-p_t},
\]
where $p_t$ is the predicted-risk threshold at which a clinician would
consider additional cardiovascular evaluation.
Because the overall 5-year MACE prevalence was 2.22\%, we focused the primary analysis on clinically plausible thresholds between 0.5\% and 10\%. For comparison,
we included the strategies of treating all patients as high risk and
treating no patients as high risk.

All models achieved positive net benefit across the full range examined. Mammo-CLIP and Mammo-FM each exceeded the treat-all reference strategy. They also exceeded the age-only floor's net benefit at every threshold examined between 0.5\% and 10\%. Tabular-only provided the highest net benefit across most of the range, closely matched by Mammo-CLIP between 3.0\% and 3.5\% (net benefit 0.0125 and 0.0115 for both models at these two thresholds, respectively). This supports the fact that information in mammograms can provide similar level of information about the patients cardiovascular health than dedicated bloodwork, without additional radiation and without the need for the radiologist's input, providing an actionable screening tool with positive net benefit.

\subsection{Association between predicted risk and radiologist-reported BAC}
\label{sec:bac}

Breast arterial calcifications (BAC) are a mammographic finding associated with cardiovascular disease. To examine whether image-based cardiovascular risk predictions captured this vascular phenotype, we assessed BAC mentions in radiology reports linked to the mammographic examinations. A patient was classified as BAC-positive if the report mentioned vascular calcification. BAC was reported in 64 of 5,287 test patients (1.21\%) and 213 of 17,210 development-cohort patients (1.24\%).

\begin{figure}[t]
  \centering
  \includegraphics[width=\columnwidth]{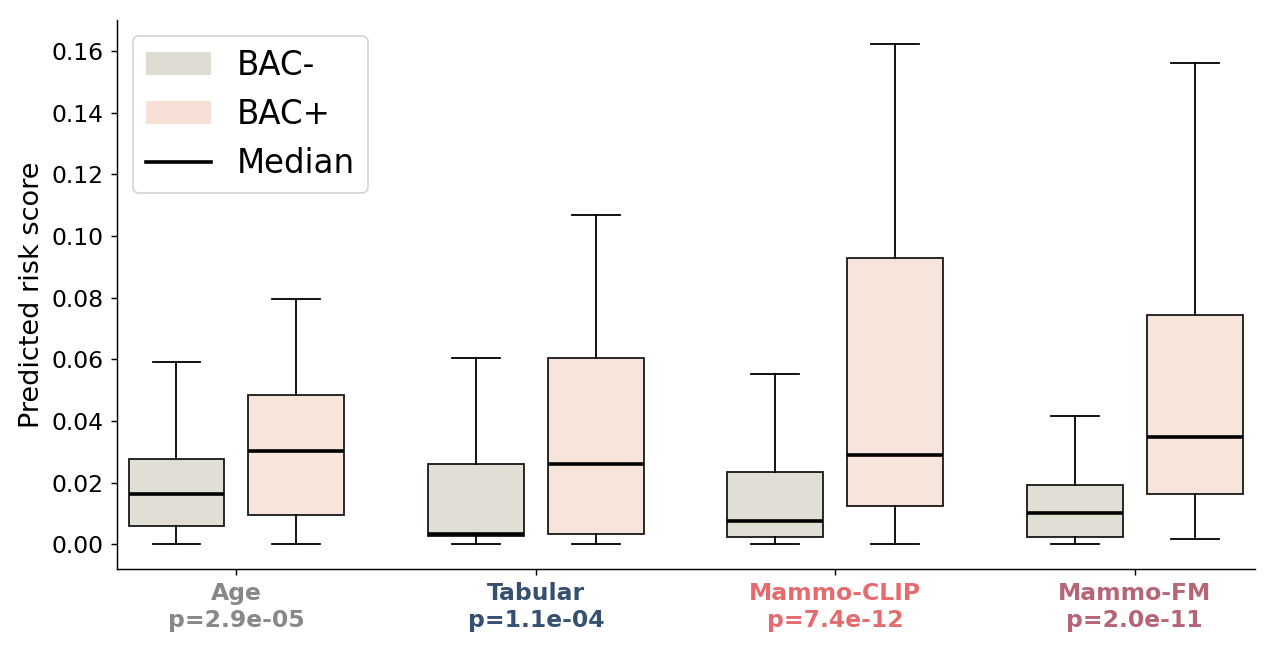}
  \caption{Predicted 5-year MACE risk among true negative
  test patients, stratified by
  radiologist reported BAC status.}
  \label{fig:bac_quintiles}
\end{figure}

Among true-negative patients (those without a qualifying 5-year MACE), BAC-positive patients had significantly higher isotonic-calibrated predicted risk across all models (Figure~\ref{fig:bac_quintiles}). The association was strongest for Mammo-CLIP (0.054 vs.\ 0.021; 2.6-fold; $p=7.4\times10^{-12}$) and Mammo-FM (0.053 vs.\ 0.021; 2.5-fold; $p=2.0\times10^{-11}$), compared with the clinical model (0.039 vs.\ 0.020; 1.9-fold; $p=1.1\times10^{-4}$) and age-only model (0.037 vs.\ 0.021; 1.7-fold; $p=2.9\times10^{-5}$).

To account for differences in age and baseline clinical risk, we additionally compared model risk percentiles after adjustment for age and, for image-based models, clinical risk. BAC positivity remained independently associated with higher predicted-risk percentiles for Mammo-FM (+13.9 points; 95\% CI, 8.6--19.2) and Mammo-CLIP (+12.7 points; 95\% CI, 7.6--17.9). In contrast, age-adjusted associations were not significant for the clinical model (+3.8 points; 95\% CI, $-1.1$--8.8; $p=0.12$) or ImageNet-pretrained encoder (+4.4 points; 95\% CI, $-1.6$--10.4; $p=0.15$).

These findings suggest that mammography-based foundation models capture a cardiovascular risk signal associated with radiologist-reported BAC that is not explained by age or baseline clinical risk alone. BAC-negative indicates the absence of a report mention, not necessarily the absence of calcifications on the mammogram.

\begin{figure*}[t]
  \centering
  \includegraphics[width=\textwidth]{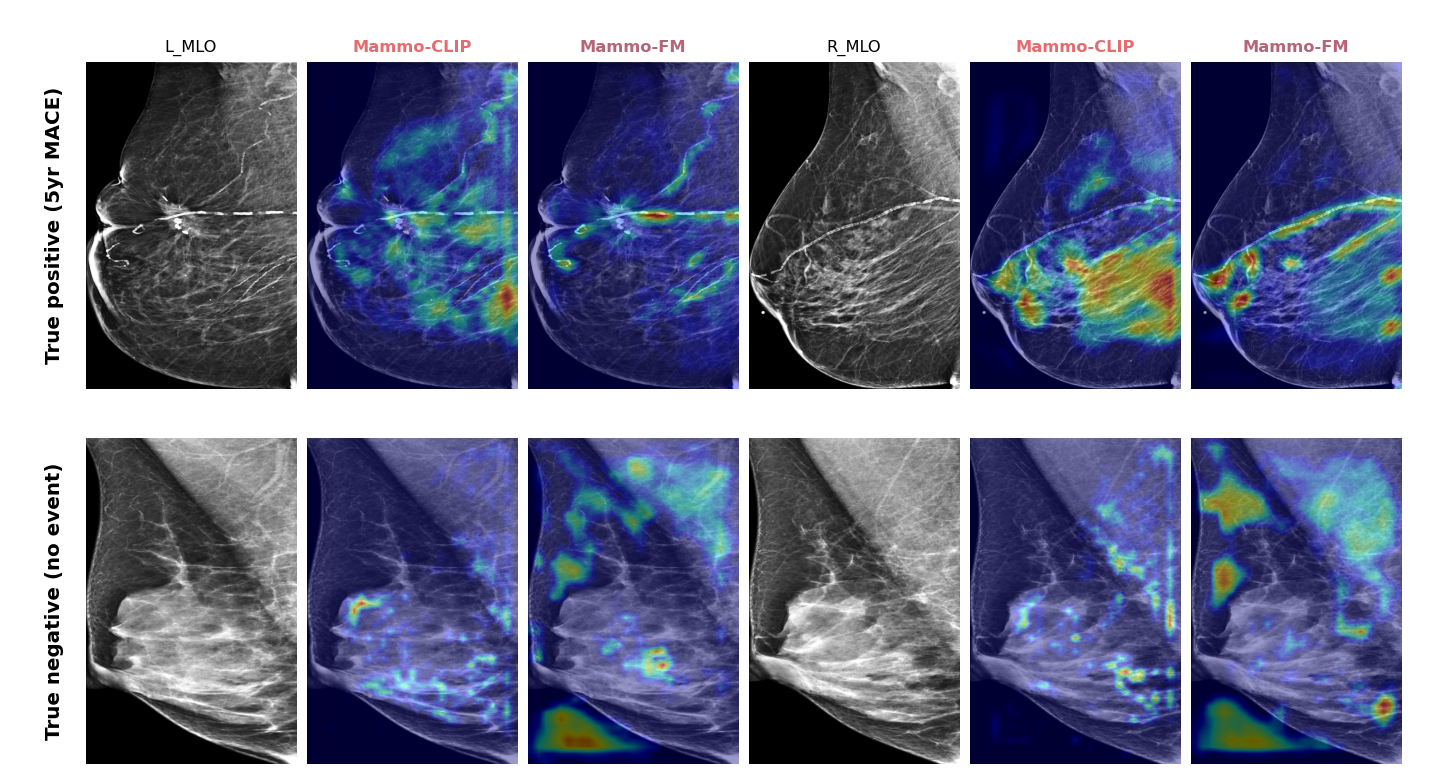}
  \caption{Qualitative gradient-based attribution of cardiovascular risk predictions. Representative cases illustrate image regions contributing to model predictions; warmer colors indicate higher positive attribution to predicted MACE risk. Attribution patterns in high risk shown examples tend to concentrate around the retroareolar region and along linear, vessel-like structures, whereas low-risk examples show more diffuse attribution across fibroglandular tissue and skin-fold regions. 
}
  \label{fig:interpretability}
\end{figure*}

\vspace{1.5em}

\subsection{Qualitative analysis of model attribution}
\label{sec:interpretability}

To qualitatively examine image regions contributing to cardiovascular risk predictions, we generated gradient-based attribution maps from the foundation-model predictors using Grad-CAM. For each mammographic view, the attribution map was projected onto the input image to visualize regions contributing to predicted MACE risk.

Attribution maps were computed using Grad-CAM at the final convolutional layer of each frozen EfficientNet-B5 backbone, with the model's predicted-risk logit as the backpropagation target. L-MLO and R-MLO views were processed independently, each producing an attribution map from a forward and backward pass through that view. Maps were overlaid on the corresponding grayscale mammogram using a jet colormap at 40\% opacity, without additional smoothing beyond Grad-CAM's native upsampling to the input resolution.

In the true-positive example (Figure~\ref{fig:interpretability}, top row), both foundation models concentrated attribution on retroareolar tissue and linear, vessel-like structures, consistent with the association between predicted risk and BAC reported in Section~\ref{sec:bac}. In contrast, attribution in the true-negative example (bottom row) was more diffuse across fibroglandular tissue and skin-fold regions, with less consistent localization to vessel-like structures. Importantly, these patterns emerged without explicit BAC detection, vessel segmentation, or quantitative measurement of calcified area. The models therefore appear capable of leveraging mammographic vascular features for cardiovascular risk prediction directly from the image, without requiring a separately engineered BAC quantification pipeline.

\section{Conclusion and Discussion}
\label{sec:discussion}

Our results show that screening mammograms contain cardiovascular risk signal that can be recovered without explicit BAC annotation or pixel-level supervision. Grad-CAM attribution maps qualitatively concentrated on vessel-like structures (Section~\ref{sec:interpretability}), particularly for Mammo-FM. Consistent with this observation, both foundation models assigned higher predicted risk to patients with radiologist-documented BAC among true negatives, despite BAC never being used as a training label (Section~\ref{sec:bac}). Together, these findings suggest that the models can leverage clinically interpretable vascular features directly from mammographic pixels rather than relying solely on nonspecific image texture.

This finding also provides a potential interpretation for some false positives. Because cardiovascular disease may remain undiagnosed or subclinical until a first clinical event, some patients identified as high-risk who did not experience a qualifying MACE during follow-up may nevertheless have underlying cardiovascular risk. This highlights the potential of opportunistic cardiovascular screening from routinely acquired screening mammograms, which could identify patients who may benefit from earlier risk assessment or preventive intervention without additional imaging.

Notably, Mammo-CLIP, pretrained on a smaller and less diverse dataset, achieved discrimination comparable to Mammo-FM. This suggests that imaging domain alignment and pretraining objectives may be at least as important as pretraining scale or dataset diversity for this task. The more spatially focused vessel-like attribution observed for Mammo-FM further suggests that similar discrimination may arise from different image representations, indicating that simply scaling pretraining data may provide limited gains without more task-aligned objectives.

Several limitations warrant further study. Our analysis uses data from a single institution, and external validation in demographically and technically distinct cohorts is needed to establish generalizability. Longitudinal analyses incorporating multiple mammograms could capture cardiovascular risk trajectories rather than a single baseline snapshot. Stratified analyses by breast density, age, and other subgroups could further characterize model performance and identify populations requiring caution. Finally, extending the fixed 5-year binary outcome to a time-to-event framework could support variable prediction horizons and better reflect potential clinical deployment.

\vspace{0.5pt}

\acks{This work was supported in part by the National Institutes of Health (NIH) through grants $1R01HL174863-01A1$ and $1U54DK144866-01$.}

\bibliography{ref}

\end{document}